\pdfoutput=1
\documentclass[a4paper,11pt, english]{article} 
\usepackage{pool/jinstpub}
\usepackage{babel}
\usepackage{graphbox}
\usepackage{multirow}
\usepackage{makecell}
\usepackage{amsmath,amssymb, gensymb}
\usepackage{natbib,hyperref}
\usepackage{url}
\usepackage{microtype}
\usepackage{subfig}
\usepackage[useregional]{datetime2}

\usepackage{siunitx}
\usepackage{xspace}

\newcommand{\NEW}{NEXT-White}

\newcommand{\tab}{table}
\newcommand{\Fig}{Figure}
\newcommand{\Eq}{Equation}

\newcommand{\micro}{\ensuremath{\mu}}

\newcommand{\bbonu}{\ensuremath{\beta\beta0\nu}}

\newcommand{\st}{\ensuremath{S_2}}
\newcommand{\so}{\ensuremath{S_1}}

\newcommand{\Qbb}{\ensuremath{Q_{\beta\beta}}}

\newcommand{\CS}{\ensuremath{^{137}}Cs}

\newcommand{\KR}{\ensuremath{^{83m}\mathrm{Kr}}\xspace}

\newcommand{\TL}{\ensuremath{{}^{208}\rm{Tl}}}

\newcommand{\THO}{\ensuremath{{}^{232}{\rm Th}}}

\newcommand{\Kr}[1]{\ensuremath{^{#1}\mathrm{Kr}}\xspace}

\DeclareSIUnit\c{\mbox{$c$}}
\DeclareSIUnit\magn{\mbox{$\times$}}
\DeclareSIUnit\min{min}
\DeclareSIUnit\week{week}
\DeclareSIUnit\year{yr}
\DeclareSIUnit\years{years}
\DeclareSIUnit\yr{yr}
\DeclareSIUnit\standard{std}
\DeclareSIUnit\str{sr}
\DeclareSIUnit\ppm{ppm}
\DeclareSIUnit\ppb{ppb}
\DeclareSIUnit\ppt{ppt}
\DeclareSIUnit\pe{PE}
\DeclareSIUnit\spe{SPE}
\DeclareSIUnit\ev{events}
\DeclareSIUnit\ct{counts}
\DeclareSIUnit\neutron{\mbox{$n$}}
\DeclareSIUnit\smp{samples}
\DeclareSIUnit\Sample{S}
\DeclareSIUnit\ch{ch}
\DeclareSIUnit\hit{hit}
\DeclareSIUnit\hits{hits}
\DeclareSIUnit\bin{(\mbox{5-PE}~bin)}
\DeclareSIUnit\sgm{\mbox{$\sigma$}}
\DeclareSIUnit\rms{RMS}
\DeclareSIUnit\keVr{\mbox{keV$_{\rm nr}$}}
\DeclareSIUnit\keVee{\mbox{keV$_{e{\rm e}}$}}
\DeclareSIUnit\ph{photon}
\DeclareSIUnit\pes{pes}
\DeclareSIUnit\el{electrons}
\DeclareSIUnit\pm{PMT}
\DeclareSIUnit\inch{"}
\DeclareSIUnit\bit{bit}
\DeclareSIUnit\sample{samples}
\DeclareSIUnit\barn{barn}
\DeclareSIUnit\bara{bar}
\DeclareSIUnit\barg{barg}
\DeclareSIUnit\mlardepth{\mbox(meter~of~\LAr~depth)}
\DeclareSIUnit\Curie{Ci}
\DeclareSIUnit\psi{psi}
\DeclareSIUnit\parsec{pc}
\DeclareSIUnit\liveday{\mbox{live-days}}
\DeclareSIUnit\days{\mbox{days}}
\DeclareSIUnit\day{\mbox{day}}
\DeclareSIUnit\miles{\mbox{miles}}
\DeclareSIUnit\degreeC{\mbox{$^{\circ}$C}}
\DeclareSIUnit\electron{\mbox{$e^-$}}
\DeclareSIUnit\Euro{\mbox{\euro}}
\DeclareSIUnit\cph{cph}
\DeclareSIUnit\neq{neq}
\DeclareSIUnit\unit{unit}
\DeclareSIUnit\byte{Byte}
\DeclareSIUnit\Bq{\becquerel}

\newcommand{\XeWaveLength}{\SI{172}{\nano\meter}}

\newcommand{\HPXe}{HPXe}
\newcommand{\HPXeEL}{HPXe-EL}

\newcommand{\RII}{Run II}

\newcommand{\XenonXRaysAverageEnergy}{\SI{30}{\keV}}
\newcommand{\XenonXRaysLineAlphaEnergy}{\SI{29.7}{\keV}}
\newcommand{\XenonXRaysLineAlphaWeightedEnergy}{\SI{29.669}{\keV}}
\newcommand{\XenonXRaysLineBetaEnergy}{\SI{34}{\keV}}
\newcommand{\XenonKAlpha}{\ensuremath{K_{\alpha}}}

\newcommand{\SQRE}{\ensuremath{1/\sqrt{E}}}

\newcommand{\Z}{\ensuremath{z}}

\newcommand{\XY}{\ensuremath{(x, y)}}

\newcommand{\XYZ}{\ensuremath{(x, y, z)}}

\newcommand{\KrEnergy}{\SI{41.5}{\keV}}

\newcommand{\TlDoubleEscapePeakEnergy}{\SI{1592.5}{\keV}}
\newcommand{\TlGammaEnergy}{\SI{2615}{\keV}}
\newcommand{\CsGammaEnergy}{\SI{661.6}{\keV}}
\newcommand{\ElectronPositronPair}{\SI{511}{\keV}}

\newcommand{\XraypeakRegion}{\ensuremath{\in (28, 32)} keV}
\newcommand{\CsPhotopeakRegion}{\ensuremath{\in (650,675)} keV}
\newcommand{\DoubleEscapeRegion}{\ensuremath{\in (1550,1640)} keV}
\newcommand{\FFZ}{\ensuremath{50~\mathrm{mm} < Z_{\mathrm{min}},\ Z_{\mathrm{max}} < 500~\mathrm{mm}}}
\newcommand{\FFR}{\ensuremath{R_{\mathrm{max}} < 180~\mathrm{mm}}}

\newcommand{\XRZ}{\ensuremath{160~\mathrm{mm} < Z_{\mathrm{min}},\ Z_{\mathrm{max}} < 300~\mathrm{mm}}}

\newcommand{\CSR}{\ensuremath{R_{\mathrm{max}} < 150~\mathrm{mm}}}

\newcommand{\ResolutionXRays}{\SI{5.71 +- 0.4}{\percent}}

\newcommand{\ResolutionCsRF}{\SI{1.45 +- 0.1}{\percent}}
\newcommand{\ResolutionTlRF}{\SI{1.11 +- 0.1}{\percent}}

\newcommand{\XenonIntrinsicEnergyResolution}{\SI{0.3}{\percent}}

\newcommand{\RunFourSevenThreeFourDate}{\DTMdisplaydate{2017}{10}{10}{-1}}
\newcommand{\RunFourSevenThreeFourType}{Kr}
\newcommand{\RunFourSevenThreeFourTriggers}{\num{2687860}}
\newcommand{\RunFourSevenThreeFourDuration}{\SI{72}{\hour}}
\newcommand{\RunFourSevenThreeFourTriggerRate}{\SI{10.5}{\hertz}}
\newcommand{\RunFourSevenThreeFourAverageLifetime}{\SI{1776}{\micro\second}}

\newcommand{\RunFourSevenThreeFiveType}{Cs/Th}
\newcommand{\RunFourSevenThreeFiveDuration}{\SI{48.4}{\hour}}
\newcommand{\RunFourSevenThreeFiveTriggerRate}{\SI{1.83}{\hertz}}
\newcommand{\RunFourSevenThreeFiveTriggers}{\num{320 039}}

\newcommand{\RunFourSevenThreeSixType}{Kr}
\newcommand{\RunFourSevenThreeSixDuration}{\SI{2.8}{\hour}}
\newcommand{\RunFourSevenThreeSixTriggerRate}{\SI{10.4}{\hertz}}
\newcommand{\RunFourSevenThreeSixTriggers}{\num{106 182}}
\newcommand{\RunFourSevenThreeSixAverageLifetime}{\SI{1805}{\micro\second}}

\newcommand{\RunFourSevenThreeSevenType}{Cs/Th}
\newcommand{\RunFourSevenThreeSevenDuration}{\SI{48.1}{\hour}}
\newcommand{\RunFourSevenThreeSevenTriggerRate}{\SI{1.84}{\hertz}}
\newcommand{\RunFourSevenThreeSevenTriggers}{\num{320 546}}

\newcommand{\RunFourSevenThreeEightType}{Kr}
\newcommand{\RunFourSevenThreeEightDuration}{\SI{3.6}{\hour}}
\newcommand{\RunFourSevenThreeEightTriggerRate}{\SI{10.4}{\hertz}}
\newcommand{\RunFourSevenThreeEightTriggers}{\num{132 751}}
\newcommand{\RunFourSevenThreeEightAverageLifetime}{\SI{1820}{\micro\second}}

\newcommand{\RunFourSevenThreeNineType}{Cs/Th}
\newcommand{\RunFourSevenThreeNineDuration}{\SI{45.4}{\hour}}
\newcommand{\RunFourSevenThreeNineTriggerRate}{\SI{1.85}{\hertz}}
\newcommand{\RunFourSevenThreeNineTriggers}{\num{302 961}}

\newcommand{\NewSevenBarPressureRunII}{\SI{7.2}{\bar}}

\newcommand{\NewPressureVesselMaterial}{316Ti}

\newcommand{\NewTpcLength}{\SI{664.5}{\mm}}

\newcommand{\NewTpcDriftLength}{\SI{530.3 +- 2}{\mm}}

\newcommand{\NewTpcELGap}{\SI{6}{\mm}}

\newcommand{\NewNumberOfSiPM}{\num{1792}}

\newcommand{\NewSipmPitch}{\SI{10}{\mm}}

\newcommand{\NewSiPMSamplingRebinned}{\SI{2}{\micro\second}}

\newcommand{\NewNumberOfPMT}{\num{12}}

\newcommand{\NewCathodeToPMTs}{\SI{130}{\mm}}

\newcommand{\NewPMTSampling}{\SI{25}{\nano\second}}

\newcommand{\NewTpcDiameter}{\SI{454}{\mm}}

\newcommand{\NewTypePMT}{Hamamatsu R11410-10}
\newcommand{\NewPMTCoverage}{31\%}

\newcommand{\NewBarrelICS}{\SI{60}{\mm}}

\newcommand{\NewPlatesICS}{\SI{120}{\mm}}

\begin{document}
\title{Initial results on energy resolution of the \NEW\ detector}
\collaboration{The NEXT Collaboration}
\author[b, 1]{J.~Renner,}

\author[j, u]{P.~Ferrario,}

\author[b, l]{G.~Mart\'inez-Lema,}

\author[p, j]{F.~Monrabal,}

\author[g]{A.~Para,}

\author[j, u, 2]{J.J.~G\'omez-Cadenas,}

\author[a]{C.~Adams,}
\author[b]{V.~\'Alvarez,}
\author[c]{L.~Arazi,}
\author[d]{C.D.R.~Azevedo,}
\author[n]{K. Bailey,}
\author[h]{F.~Ballester,}
\author[b]{J.M.~Benlloch-Rodr\'{i}guez,}
\author[e]{F.I.G.M.~Borges,}
\author[b]{A.~Botas,}
\author[b]{S.~C\'arcel,}
\author[b]{J.V.~Carri\'on,}
\author[f]{S.~Cebri\'an,}
\author[e]{C.A.N.~Conde,}
\author[b]{J.~D\'iaz,}
\author[g]{M.~Diesburg,}
\author[e]{J.~Escada,}
\author[h]{R.~Esteve,}
\author[b]{R.~Felkai,}
\author[i]{A.F.M.~Fernandes,}
\author[i]{L.M.P.~Fernandes,}
\author[d]{A.L.~Ferreira,}
\author[i]{E.D.C.~Freitas,}
\author[j]{J.~Generowicz,}

\author[k]{A.~Goldschmidt,}
\author[l]{D.~Gonz\'alez-D\'iaz,}
\author[a]{R.~Guenette,}
\author[m]{R.M.~Guti\'errez,}
\author[n]{K.~Hafidi,}
\author[o]{J.~Hauptman,}
\author[i]{C.A.O.~Henriques,}
\author[m]{A.I.~Hernandez,}
\author[l]{J.A.~Hernando~Morata,}
\author[h]{V.~Herrero,}
\author[n]{S.~Johnston,}
\author[p]{B.J.P.~Jones,}
\author[b]{M.Kekic,}
\author[q]{L.~Labarga,}
\author[b]{A.~Laing,}
\author[g]{P.~Lebrun,}
%
%
\author[b]{N.~L\'opez-March,}
\author[m]{M.~Losada,}
\author[i]{R.D.P.~Mano,}
\author[a]{J.~Mart\'in-Albo,}
\author[b]{A.~Mart\'inez,}
\author[p]{A.D.~McDonald,}
\author[i]{C.M.B.~Monteiro,}
\author[h]{F.J.~Mora,}
%
%
\author[b]{J.~Mu\~noz Vidal,}
\author[b]{M.~Musti,}
\author[b]{M.~Nebot-Guinot,}
\author[b]{P.~Novella,}
\author[p,2]{D.R.~Nygren,}
\author[b]{B.~Palmeiro,}
\author[b,3]{J.~P\'{e}rez,}
\author[p]{F.~Psihas,}
\author[b]{M.~Querol,}
\author[n]{J.~Repond,}
\author[n]{S.~Riordan,}
\author[r]{L.~Ripoll,}
\author[b]{J.~Rodr\'iguez,}
\author[p]{L.~Rogers,}

\author[b]{C.~Romo-Luque,}

\author[e]{F.P.~Santos,}
\author[i]{J.M.F.~dos~Santos,}
\author[b]{A.~Sim\'on,}
\author[s,4]{C.~Sofka,}
\author[b]{M.~Sorel,}
\author[s]{T.~Stiegler,}
\author[h]{J.F.~Toledo,}
\author[j]{J.~Torrent,}
%
%
\author[d]{J.F.C.A.~Veloso,}
\author[s]{R.~Webb,}
\author[s,3]{J.T.~White,}
\author[b]{N.~Yahlali}

\note{corresponding author.}
\note{NEXT Co-spokesperson.}
\note{Now at Laboratorio Subterr\'{a}neo de Canfranc, Spain.}
\note{Now at University of Texas at Austin, USA.}
\note{Deceased.}

\emailAdd{josren@uv.es}
\affiliation[a]{
  Department of Physics, Harvard University\\
  Cambridge, MA 02138, USA}
\affiliation[b]{
Instituto de F\'isica Corpuscular (IFIC), CSIC \& Universitat de Val\`encia\\
Calle Catedr\'atico Jos\'e Beltr\'an, 2, 46980 Paterna, Valencia, Spain}
\affiliation[c]{
  Nuclear Engineering Unit, Faculty of Engineering Sciences, Ben-Gurion University of the Negev\\
P.O.B. 653 Beer-Sheva 8410501, Israel}
\affiliation[d]{
Institute of Nanostructures, Nanomodelling and Nanofabrication (i3N), Universidade de Aveiro\\
Campus de Santiago, 3810-193 Aveiro, Portugal}
\affiliation[e]{
LIP, Department of Physics, University of Coimbra\\
P-3004 516 Coimbra, Portugal}
\affiliation[f]{
Laboratorio de F\'isica Nuclear y Astropart\'iculas, Universidad de Zaragoza\\ 
Calle Pedro Cerbuna, 12, 50009 Zaragoza, Spain}
\affiliation[g]{
Fermi National Accelerator Laboratory\\ 
Batavia, Illinois 60510, USA}
\affiliation[h]{
Instituto de Instrumentaci\'on para Imagen Molecular (I3M), Centro Mixto CSIC - Universitat Polit\`ecnica de Val\`encia\\
Camino de Vera s/n, 46022 Valencia, Spain}
\affiliation[i]{
LIBPhys, Physics Department, University of Coimbra\\
Rua Larga, 3004-516 Coimbra, Portugal}
\affiliation[j]{
Donostia International Physics Center (DIPC)\\
Paseo Manuel Lardizabal 4, 20018 Donostia-San Sebastian, Spain}
\affiliation[k]{
Lawrence Berkeley National Laboratory (LBNL)\\
1 Cyclotron Road, Berkeley, California 94720, USA}
\affiliation[l]{
Instituto Gallego de F\'isica de Altas Energ\'ias, Univ.\ de Santiago de Compostela\\
Campus sur, R\'ua Xos\'e Mar\'ia Su\'arez N\'u\~nez, s/n, 15782 Santiago de Compostela, Spain}
\affiliation[m]
{Centro de Investigaci\'on en Ciencias B\'asicas y Aplicadas, Universidad Antonio Nari\~no\\ 
Sede Circunvalar, Carretera 3 Este No.\ 47 A-15, Bogot\'a, Colombia}
\affiliation[n]
{Argonne National Laboratory,\\ 
Argonne IL 60439, USA}
\affiliation[o]{
Department of Physics and Astronomy, Iowa State University\\
12 Physics Hall, Ames, Iowa 50011-3160, USA}
\affiliation[p]{
Department of Physics, University of Texas at Arlington\\
Arlington, Texas 76019, USA}
\affiliation[q]{
Departamento de F\'isica Te\'orica, Universidad Aut\'onoma de Madrid\\
Campus de Cantoblanco, 28049 Madrid, Spain}
\affiliation[r]{
Escola Polit\`ecnica Superior, Universitat de Girona\\
Av.~Montilivi, s/n, 17071 Girona, Spain}
\affiliation[s]{
Department of Physics and Astronomy, Texas A\&M University\\
College Station, Texas 77843-4242, USA}
%
%
\affiliation[u]{
IKERBASQUE, Basque Foundation for Science\\
48013 Bilbao, Spain.}

\abstract{One of the major goals of the \NEW\ (NEW) detector is to demonstrate the energy resolution that an electroluminescent high pressure xenon TPC can achieve for high energy tracks. For this purpose, energy calibrations with \CS\ and \THO\ sources have been carried out as a part of the long run taken with the detector during most of 2017. This paper describes the initial results obtained with those calibrations, showing excellent linearity and an energy resolution that extrapolates to approximately 1\% FWHM at \Qbb.}

\arxivnumber{1808.01804}

\collaboration{\includegraphics[height=9mm]{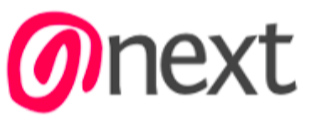}\\[6pt]
	NEXT collaboration}

\maketitle
\flushbottom
	
\section{Introduction}
 Excellent energy resolution is one of the most important tools to reject backgrounds in neutrinoless double beta decay (\bbonu) searches. The low value of the Fano factor in high-pressure xenon \cite{BOLOTNIKOV1997360} offers the possibility to build a gaseous time projection chamber (TPC) with ultimate energy resolution of \XenonIntrinsicEnergyResolution\ FWHM at \Qbb = 2457.8 keV, provided that proportional amplification such as that granted by electroluminescence is used \cite{Nygren:2009zz} and all other sources of systematic errors are under control. The NEXT experimental program \cite{Alvarez:2011my, Alvarez:2012haa, Gomez-Cadenas:2013lta, Martin-Albo:2015rhw} is exploiting this opportunity and it is pioneering the development of electroluminescent high pressure xenon chambers (\HPXeEL) for \bbonu\ searches aiming at the 100 kg-class, NEXT-100 detector.
 
 Small scale detectors developed at LBNL and at IFIC have demonstrated the excellent energy resolution of the electroluminescent high-pressure Xe TPC \cite{Alvarez:2012hh, Alvarez:2012xda}, consistent with the expectations. An intermediate scale detector \NEW\footnote{Named after Prof. James White, our late mentor and friend.} has been constructed at IFIC and it is currently operating at the Canfranc Underground Laboratory (LSC)  \cite{Monrabal:2018xlr}. Its calibration run with a \Kr{83m} source \cite{Martinez-Lema:2018ibw} has demonstrated that very good energy resolution for low-energy/pointlike \KrEnergy\ depositions can be attained in larger size detectors as well.
 
 In this work, we report studies of the energy resolution obtained from measurements using \CS\ and \THO\ calibration sources at a pressure of  \NewSevenBarPressureRunII, during the last part of the so-called \RII, which extended through Fall 2017. These sources provide monoenergetic events at higher energies and thus they offer stringent tests of the energy reconstruction in the conditions similar to the expected double beta decay signals. 

The organization of this paper is as follows: section \ref{sec.setup} describes the experimental setup; the event selection and reconstruction is presented in section \ref{sec.selection};  section \ref{sec.ecal} discusses the absolute energy calibration; section \ref{sec.eres} presents the results on the energy resolution over wide energy range. Summary and discussion is presented in section \ref{sec.conclu}.
\section{The \NEW ~ detector}
\label{sec.setup}

\begin{figure}[tbh]
\centering
\includegraphics[width= 0.7\textwidth]{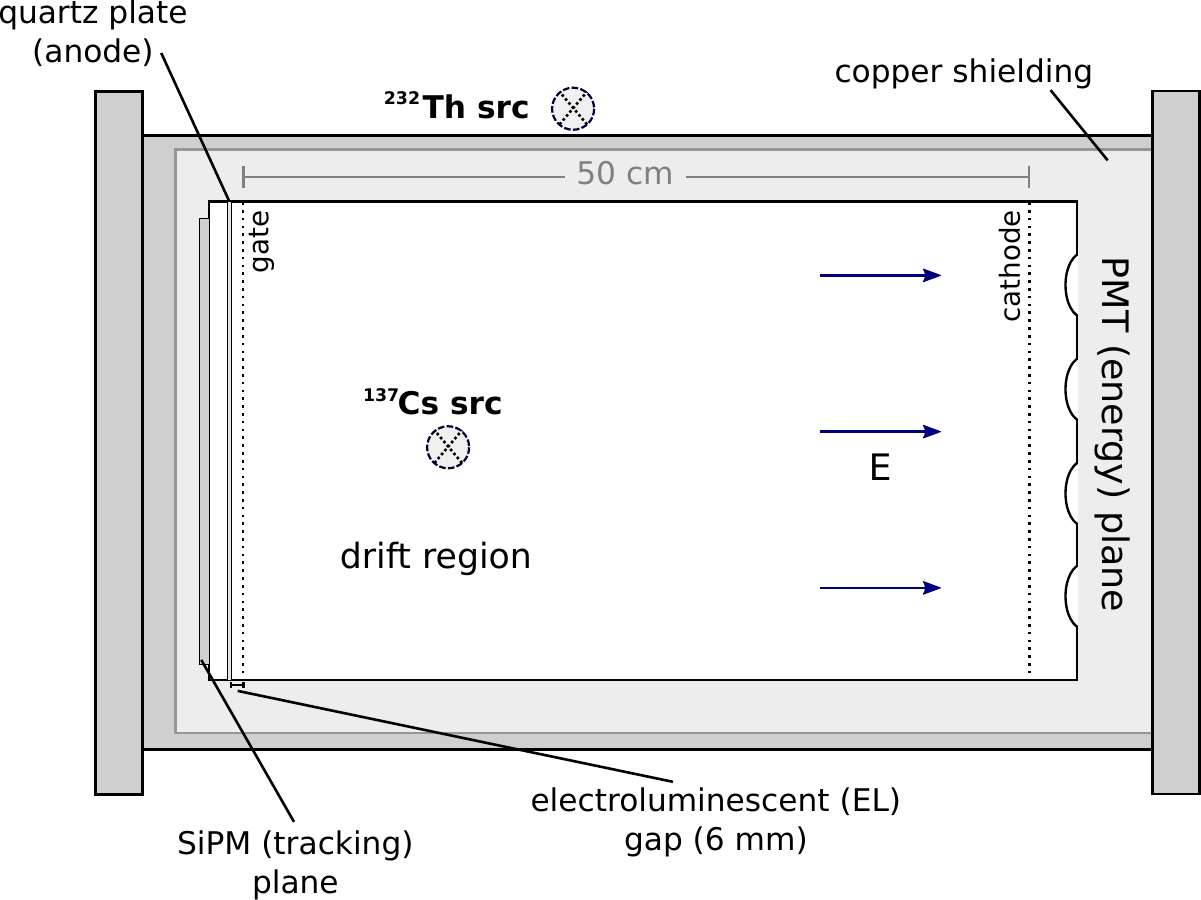}
\caption{Schematic of the detector and source configuration in this study (not drawn to scale).  The \CS\ source was placed at the lateral entrance port of the pressure vessel, and the \THO\ source was placed on top of the pressure vessel.}
\label{fig.config}
\end{figure}

\begin{table}[htb]
\caption{\NEW  ~TPC operational parameters.}
\begin{center}
\begin{tabular}{|c|c|}
\hline
TPC parameter & Value \\
\hline
Pressure & 7.2 bar \\
$E/p$ & 1.7 kV cm$^{-1}$ bar$^{-1}$ \\
Drift field & 400 V cm$^{-1}$ \\
$V_{\mathrm{cathode}}$ & -28 kV \\
$V_{\mathrm{gate}}$ & -7.0 kV \\
Length & \NewTpcLength \\
Diameter &  \NewTpcDiameter \\
EL gap & \NewTpcELGap \\
Drift length & \NewTpcDriftLength \\
Fiducial mass & 2.3 kg \\
\hline\hline
\end{tabular}
\end{center}
\label{tab.TPC}
\end{table}%

The \NEW\ apparatus has been described with great detail elsewhere \cite{Martinez-Lema:2018ibw}. The main subsystems of the detector are the TPC, the energy plane and the tracking plane. The operating parameters of the TPC  used in this study are described in table \ref{tab.TPC}. The energy plane is instrumented with \NewNumberOfPMT\ \NewTypePMT\  PMTs located \NewCathodeToPMTs\ behind the cathode, covering \NewPMTCoverage\ of the end-plate area. The tracking plane is instrumented with \NewNumberOfSiPM\ SensL series-C silicon photomultipliers (SiPMs) arranged in a grid with a pitch of \NewSipmPitch. An ultra-pure inner copper shell (ICS) \NewBarrelICS\ thick acts as a shield in the barrel region. The tracking plane and the energy plane are supported and shielded by pure copper plates \NewPlatesICS\ thick.  

The detector is enclosed in a pressure vessel fabricated from \NewPressureVesselMaterial, a titanium-stabilized alloy of stainless steel, and placed on a seismic table within a lead shield that can be opened and closed mechanically.  To ensure long electron lifetime, the xenon is constantly circulated through a gas system containing a hot getter to remove impurities. The detector, gas system, and readout electronics are all elevated above the ground on a tramex platform in HALL-A of the Laboratorio Subterr\'{a}neo de Canfranc (LSC).

As a charged particle propagates in the dense gas of \NEW\ it loses energy by ionization and excitation of atoms of the medium. The excited atoms return to the ground state by a prompt emission of VUV (\XeWaveLength)  scintillation light (\so).  Ionization electrons drift toward the TPC anode where they produce an amplified light signal (\st) inside the electroluminescent region composed of a transparent mesh, the gate, and a quartz plate coated with conductive indium tin oxide (ITO), the anode. The \so\ and \st\ signals are recorded by the PMTs of the energy plane. The \st\ signal is used to trigger the data acquisition and  to determine the total energy deposition of the event. The time difference between \so\ and \st\ signals  provides the timing information used to localize the event within the drift volume. The \st\ signal is also recorded by the dense grid of SiPMs, the tracking plane, located in close proximity to the anode. The spatial distribution of signals observed in the tracking plane  yields  the transverse position of the arriving ionization electron with a precision of a few millimeters.

\begin{figure}[htb]
	\centering
	\includegraphics[width= 1.0\textwidth]{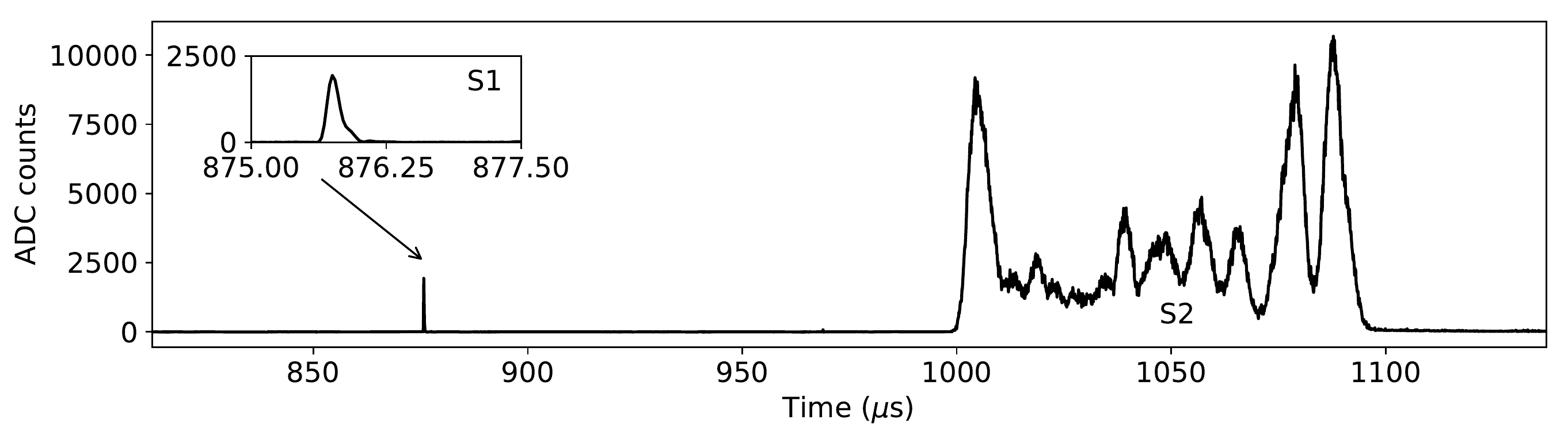}
	\caption{An example of a waveform recorded by the PMTs for an event with an approximate total energy deposition of 1.6 MeV.}\label{fig.waveform}
\end{figure}

\Fig\ \ref{fig.waveform} shows an example of a waveform recorded by the PMTs of the energy plane for an event with an approximate total energy deposition of 1.6 MeV. The prompt primary scintillation signal \so\ appears near the beginning of the buffer, while the EL signal, \st, extends for tens of $\mu$s.  The time elapsed between detection of the \so\ signal and detection of each of the components of the \st\ signal  reflects the drift time of the ionization electrons. Its measurement, together with the known value of the drift velocity (approximately 1 mm/$\mu$s \cite{Simon:2018vep}), determines the z-coordinate at which each of the ionization electrons was produced in the active region. The \XY\ coordinate is obtained by a position reconstruction algorithm which uses the energies recorded by the SiPMs of the tracking plane.  The combination of the PMT and SiPM sensor responses yields a full reconstruction of 3D spatial position and an energy measurement of the arriving ionization electrons.

The detector response depends on the spatial position of the original energy deposition: finite electron lifetime is the origin of the z-dependence of the response, whereas the local variation of the electroluminesce gain and light collection results in the variation of detector response with the transverse x-y position. Detailed response maps of the detector were determined from the dedicated calibration runs using \KR source events distributed uniformly throughout the detector volume \cite{Martinez-Lema:2018ibw}. The low energy \KR\ events represent nearly pointlike energy depositions in the detector volume. While this is an optimal choice for the determination of the local response maps, it provides little insight into the energy resolution for higher-energy, extended events.

Studies of the response linearity and energy resolution of the \NEW\ detector at high energies have been carried out using dedicated runs where the external radioactive sources were used as a source of monoenergetic, yet spatially extended, energy depositions. Two sources were employed simultaneously for this purpose: \CS\ and \THO. The \CS\ source was placed at the lateral port of the pressure vessel, and the \THO\ source was placed on top of the vessel as illustrated in Figure  \ref{fig.config}.  These sources provide two well defined calibration lines: \CS\ decays by emitting a gamma ray with energy of  \CsGammaEnergy\ and \THO\ decays eventually to \TL\ which emits a gamma ray with an energy of  \TlGammaEnergy. The data acquisition was triggered by a  S2 signal in two PMTs integrated over an interval of 10-250 $\mu$s exceeding a specified trigger threshold. The trigger threshold was selected to ensure full detection efficiency for the \CS\ line and higher energy depositions.

The data sets used in this analysis were taken at the operational pressure of \NewSevenBarPressureRunII.  Prior to each data run taken with the \CS\ and \THO\ sources, a dataset was taken with the trigger set to capture 41.5 keV events produced by decays of \KR. These point-like events were used to make a precise determination of response maps and electron lifetime in the active region as described in \cite{Martinez-Lema:2018ibw}.  Run 4734 started on 
\RunFourSevenThreeFourDate, and the subsequent runs were taken over a period of two weeks, as summarized in \tab\ \ref{tbl.runs}.

\begin{table}
	\begin{center}
		\caption{Data acquisition summary.}\label{tbl.runs}
		\begin{tabular}{ccrrrr}
			Run \# & Type & Duration & Avg. Rate & Triggers & Average Lifetime\\
			\hline
			4734 &    \RunFourSevenThreeFourType   & \RunFourSevenThreeFourDuration & \RunFourSevenThreeFourTriggerRate & \RunFourSevenThreeFourTriggers & \RunFourSevenThreeFourAverageLifetime\\
			4735 & \RunFourSevenThreeFiveType &\RunFourSevenThreeFiveDuration & \RunFourSevenThreeFiveTriggerRate & \RunFourSevenThreeFiveTriggers & --- \\
			4736 &   \RunFourSevenThreeSixType    & \RunFourSevenThreeSixDuration&  \RunFourSevenThreeSixTriggerRate & \RunFourSevenThreeSixTriggers& \RunFourSevenThreeSixAverageLifetime\\
			4737 & \RunFourSevenThreeSevenType &\RunFourSevenThreeSevenDuration & \RunFourSevenThreeSevenTriggerRate & \RunFourSevenThreeSevenTriggers & --- \\
			4738 &    \RunFourSevenThreeEightType    & \RunFourSevenThreeEightDuration&  \RunFourSevenThreeEightTriggerRate & \RunFourSevenThreeEightTriggers& \RunFourSevenThreeEightAverageLifetime\\
			4739 & \RunFourSevenThreeNineType &\RunFourSevenThreeNineDuration & \RunFourSevenThreeNineTriggerRate & \RunFourSevenThreeNineTriggers & --- \\
		\end{tabular}
	\end{center}
\end{table}

\section{Selection and reconstruction of events}
\label{sec.selection}

Studies presented here focus on events produced in the region of the \CS\, photo-peak (661 keV) and in the region of the \TL\ ``double-escape'' peak (\TlDoubleEscapePeakEnergy). The latter corresponds to events in which the \TlGammaEnergy\ gamma emitted in a \TL\ decay converts in an electric field of a nucleus  into an electron-positron pair, and the two \ElectronPositronPair\ gamma rays produced by the annihilation of the positron escape undetected. 

Interactions of ionizing radiation in xenon can occasionally eject electrons from inner shells of the  xenon atoms. The subsequent transitions of electrons into these lower shells lead to the emission of characteristic X-rays, several of which have energies near \XenonXRaysAverageEnergy. Some of these  X-rays may travel a significant distance from the main ionization track before interacting, or even escape from the detector entirely.  The interacting X-rays provide an additional source of nearly monoenergetic lines which can be used for the calibration.

\subsection*{Event selection}
Event characteristics were computed from the acquired PMT and SiPM waveforms.  The \so\ and \st\ signals were identified as peaks in the summed PMT waveform (sampled in bins of width \NewPMTSampling) based on their location and duration in time (see \cite{Martinez-Lema:2018ibw} for details).  Events with exactly one \so\ peak and at least one but no more than three \st\ peaks were accepted for further analysis.  

\begin{figure}[htb]
	\centering
	\includegraphics[width= 1.0\textwidth]{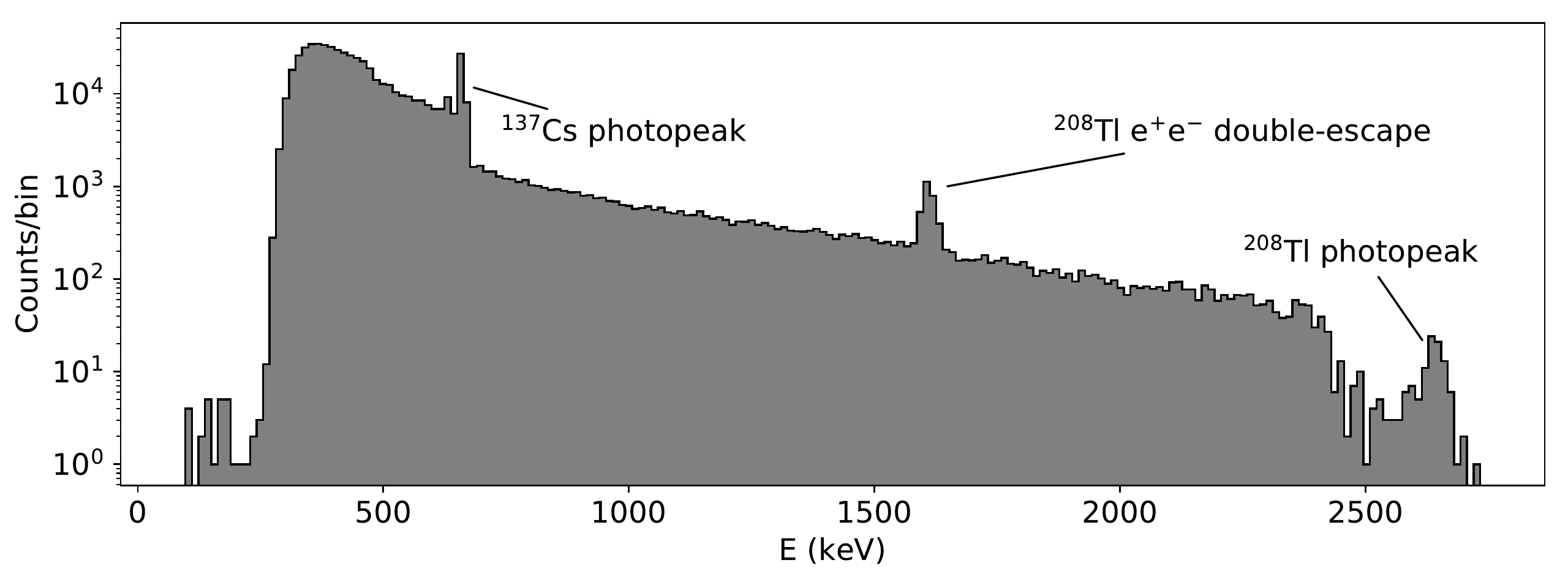}
	\caption{Energy distribution of events recorded in the full fiducial region.  The $^{137}$Cs ($\sim$662 keV) and $^{208}$Th ($\sim$2.6 MeV) photopeaks are visible along with their Compton spectra.  A peak is also visible at $\sim$1.6 MeV due to $e^{+}e^{-}$ pair production from the 2.6 MeV gamma and the escape of both 511 keV gammas produced in the resulting positron annihilation.} \label{fig.espectrum}
\end{figure}

\subsection*{Energy measurement}

The waveform of summed PMT signals was integrated in \NewSiPMSamplingRebinned\ intervals corresponding to the transit time of an electron across the EL gap.  The pattern of the observed SiPM signals in each of the time slices was used to determine the number and \XY\ positions of separate energy depositions, or "clusters", present in the given time interval. If more than one cluster was reconstructed in a given time interval, the energy detected by the PMTs was distributed proportionally amongst the clusters according to their charges, as determined by the SiPM signals. Spatial coordinates \XYZ\ of each cluster were used to determine the necessary energy correction factor for the response non-unifomity in \XY\ and the electron lifetime, as descibed in \cite{Martinez-Lema:2018ibw}. 

The total energy of the event is defined as the sum of the energies of all clusters reconstructed in the event. The resulting energy distribution of events reconstructed in the full fiducial volume is shown in \Fig\ \ref{fig.espectrum}. The full fiducial volume includes most of the active volume of the detector and is defined as \FFZ, and \FFR\ where $Z_{\mathrm{min}}$ and $Z_{\mathrm{max}}$ are the minimum and maximum z-coordinates, and $R_{\mathrm{max}}$ is the maximum radial coordinate of all reconstructed clusters in a given event.  Several peaks are visible which will be analyzed in more detail later, including the $^{137}$Cs photopeak and the $^{208}$Tl e$^{+}$e$^{-}$ double-escape peak.  The $^{208}$Tl photopeak is also visible in the data but few events were acquired, as the tracks at such energies are too long to be consistently contained within the TPC at the current operating pressure.  Because of this, a detailed analysis of this photopeak is impractical in this dataset and will be left to a future study at higher pressure.  Note that though the trigger was set to acquire events above approximately 250 keV, it was still possible to examine isolated clusters of energy deposited within higher-energy events, and thus we were also able to analyze the energy peaks corresponding to xenon characteristic x-rays (see section \ref{sec.eres}).

\subsection*{Spatial distribution of events and restricted fiducial cuts}
\label{sec.efid}

\begin{figure}[p!]
	\centering
	\includegraphics[width= 1.0\textwidth]{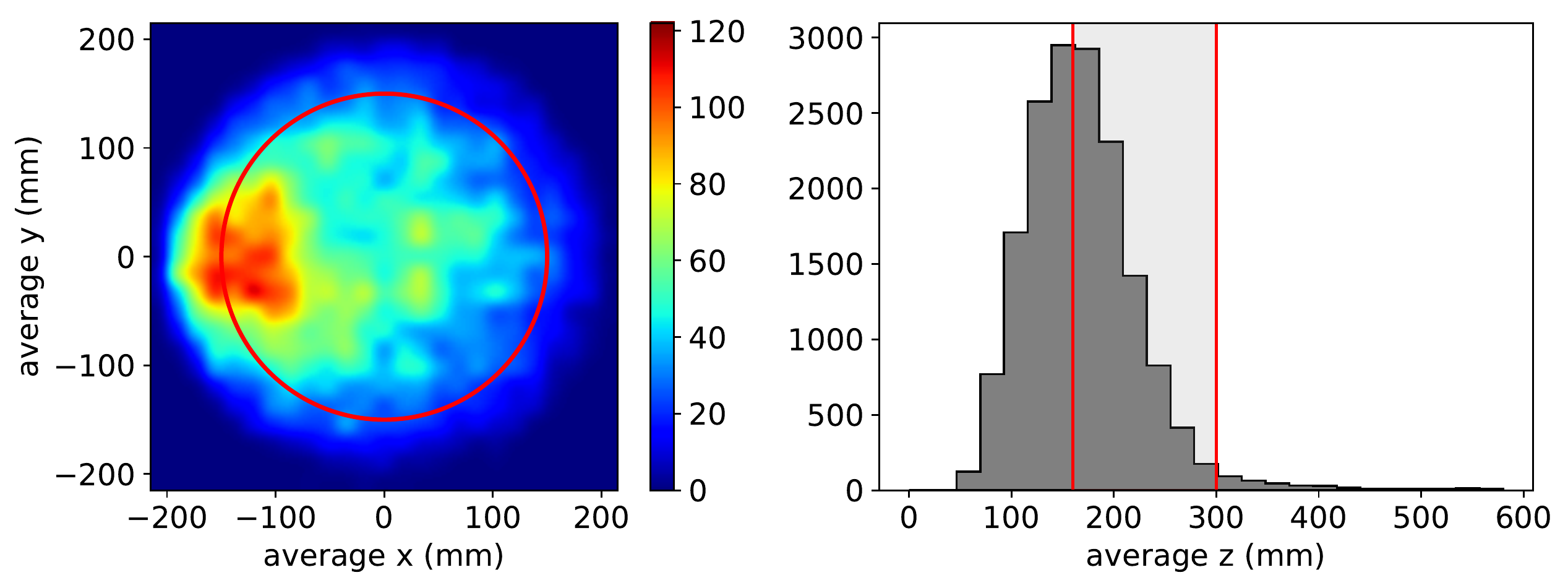}
	\includegraphics[width= 1.0\textwidth]{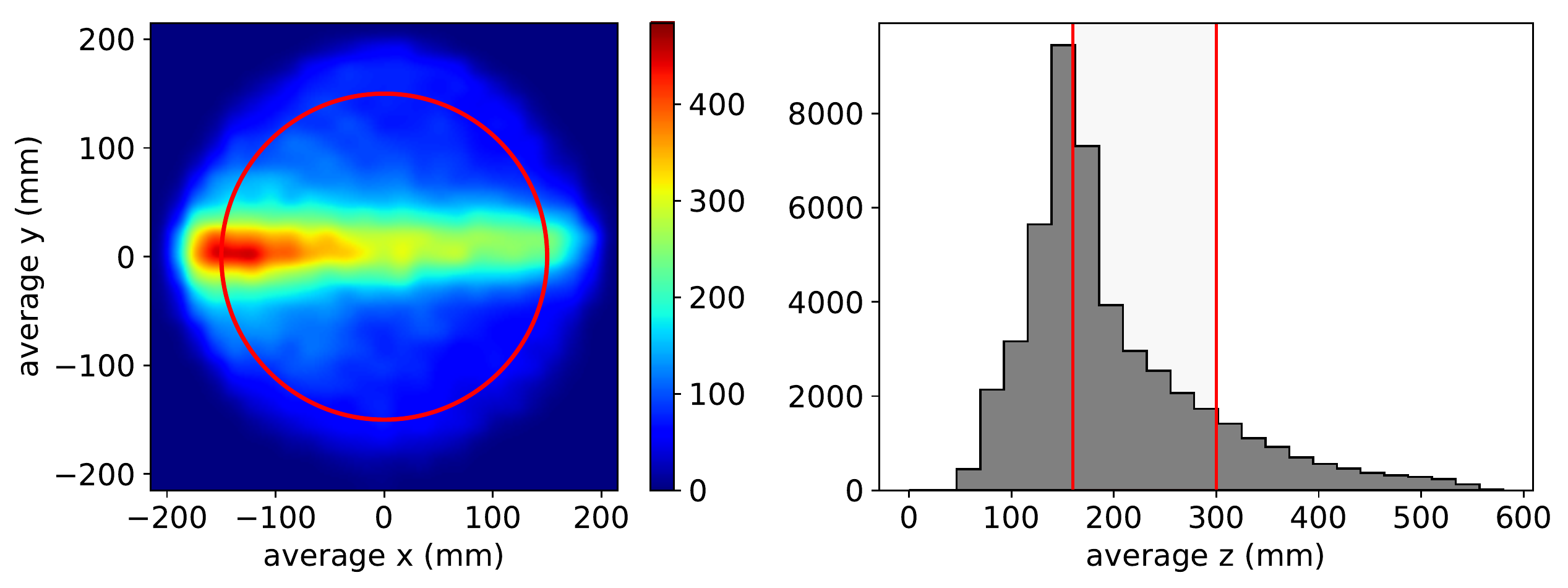}
	\includegraphics[width= 1.0\textwidth]{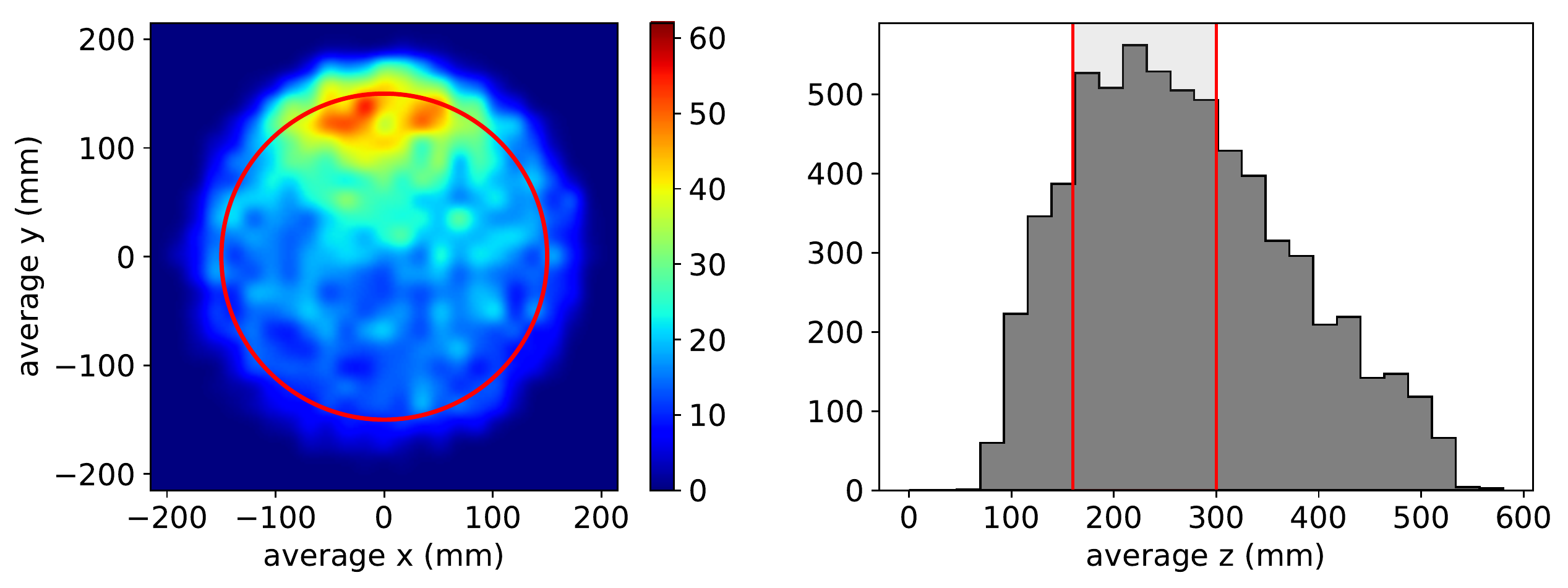}
	\caption{Distributions of average event locations in \XY\ (left) and \Z\ (right), for events in the region of the xenon x-ray peak \XraypeakRegion\ (top), the \CS\ photo-peak, \CsPhotopeakRegion\ (middle), and events in the region of the double-escape peak, \DoubleEscapeRegion\ (bottom).  The solid red lines indicate the restricted fiducial regions described in section \ref{sec.efid}.} \label{fig.xyzdist}
\end{figure}

\Fig\ \ref{fig.espectrum} indicates the presence of the calibration lines of interest, albeit with significant background. The purity of the calibration sample can be enhanced by taking advantage of the spatial localization of the parent source samples. As shown in \Fig\ \ref{fig.xyzdist} the majority of the events in the monoenergetic lines are located near the  source positions. To improve the signal-to-noise ratio of the calibration sample, a fiducial cut, \XRZ, is applied in the further analysis.

The detector response degrades significantly at the outer radii of the active volume due to imperfect light collection by the PMTs \cite{Martinez-Lema:2018ibw}, thus contributing significantly to the degradation of the energy resolution. To eliminate the bias induced by the events approaching the edges of the detector the fiducial cut \CSR\ is applied.

\section{Absolute energy calibration}
\label{sec.ecal}

\begin{figure}[htb]
	\centering
	\includegraphics[width= 1.0\textwidth]{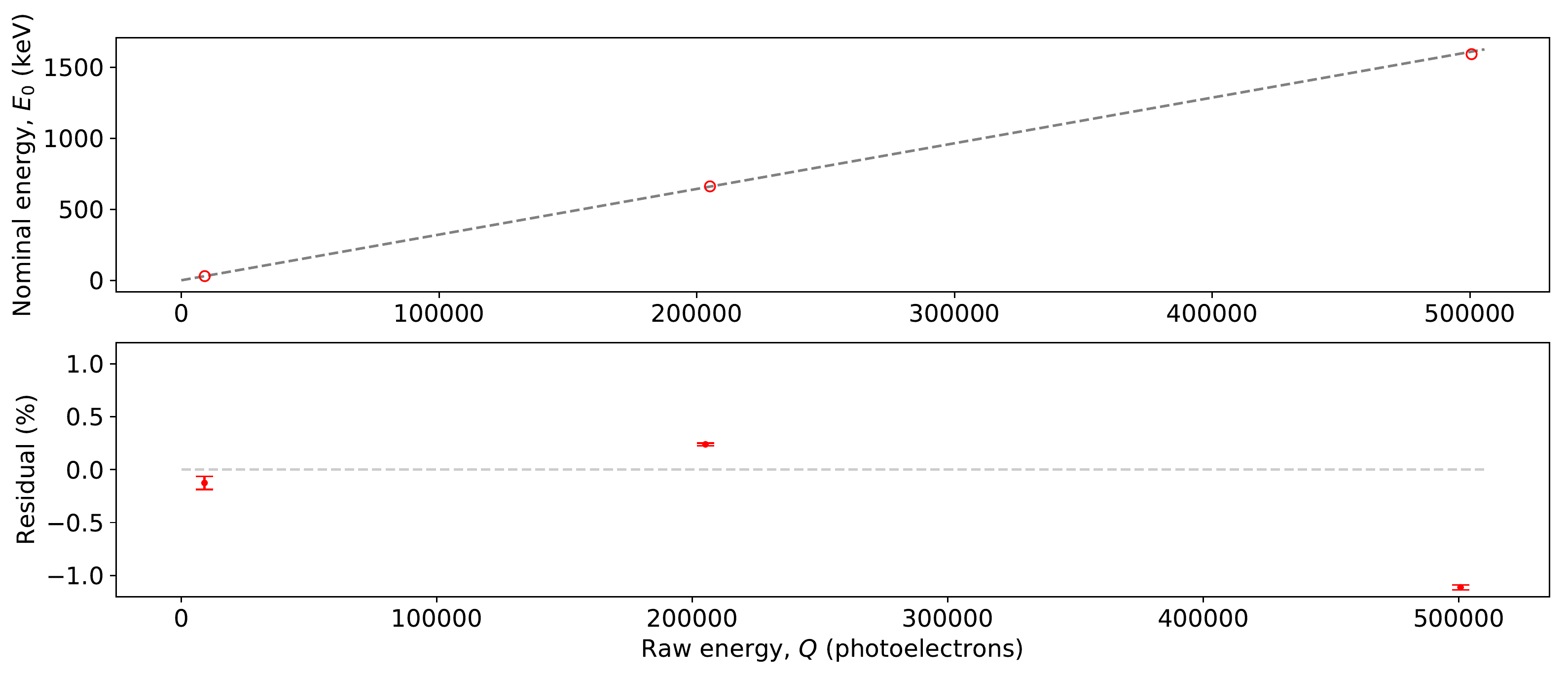}
	\caption{Energy calibration in \NEW\ resulting from a linear fit to three measured energies over a range corresponding to approximately 1.6 MeV.} \label{fig.energycalibration}
\end{figure}

The energy of the event $Q$, determined by integrating the measured \st\ signal, is expressed in photoelectrons using conversion factors determined by calibration of the PMTs.  To determine the absolute energy scale the measured detector response was compared to nominal values of the event energies. Three event samples were used for this purpose: the 29.7 keV xenon X-ray peak, the \CS\ peak, and the double-escape peak.   The nominal energies were taken from \cite{LBLTOI} and include the intensity-weighted average of the \XenonKAlpha\ x-ray lines (\XenonXRaysLineAlphaWeightedEnergy), the energy of the \CS\ emitted gamma (\CsGammaEnergy), and the energy of the double-escape peak  (\TlDoubleEscapePeakEnergy).

A linear fit was performed to obtain calibrated energy $E$~ from uncalibrated energy $Q$~ as

\begin{equation}
E = a_0\cdot Q + a_1,
\end{equation}

\noindent and the fitted parameters were found to be $a_0 = 3.2154 \pm 0.0003$ eV/photoelectron and $a_1 = 0.49975 \pm 0.01359$ keV.  The fit is shown in Fig. \ref{fig.energycalibration} along with residuals $r$ defined as the percent difference of the calibrated energy $E$~ from the nominal energy $E_0$, $r = 100\cdot(E_0 - E(Q))/E_0$, for each point $(Q,E_0)$~ used in the fit.  The errors on the measured raw energies $Q$ in the fit were taken to be the errors on the mean values of the Gaussian components of the fits to the three peaks in photoelectrons, and these errors were found to be less than $0.05\%$ of their respective means. The calibration demonstrates that the detector response is linear  within $\sim$1\% in the energy range from \XenonXRaysLineAlphaEnergy\ to \TlDoubleEscapePeakEnergy.

\section{Energy resolution}
\label{sec.eres}

From a sample of monoenergetic events the energy resolution can be determined as the ratio of FWHM of the energy distribution to the mean value of the response. At low energies the energy depositions are nearly point-like and the energy resolution is dominated by the stochastic term reflecting the fluctuations in the production of electron-ion pairs. Energy measurement for high energy events involves a sum of the energy depositions over the larger volume of the detector. They need to be corrected for the variation of the response due to the electron lifetime and/or the local inhomogeneities of the detector, and the residual imperfections of the corrections may contribute to the energy resolution (a systematic error). The relative importance of various contributions may be estimated by comparison of the observed energy resolution at different energies: the stochastic term is expected to follow \SQRE\ dependence on the energy whereas the systematic contribution is expected to be independent of the energy. Thus it is expected that the variation of the energy resolution with energy should follow

\begin{equation}
R(E) = \sqrt{\bigl(a/\sqrt{E}\bigr)^2 + c^2}
\end{equation}

\noindent or equivalently

\begin{equation}\label{eq.resolution}
R^2(E) = a^2/E + c^2
\end{equation}

\noindent where $a/\sqrt{E}$ is the stochastic term and $c$ is the systematic, constant term. Potential noise contributions, behaving like $b/E$, are expected to be negligible in our case.

\subsection*{Xenon X-rays}

\begin{figure}[htb]
	\centering
	\includegraphics[width= 1.0\textwidth]{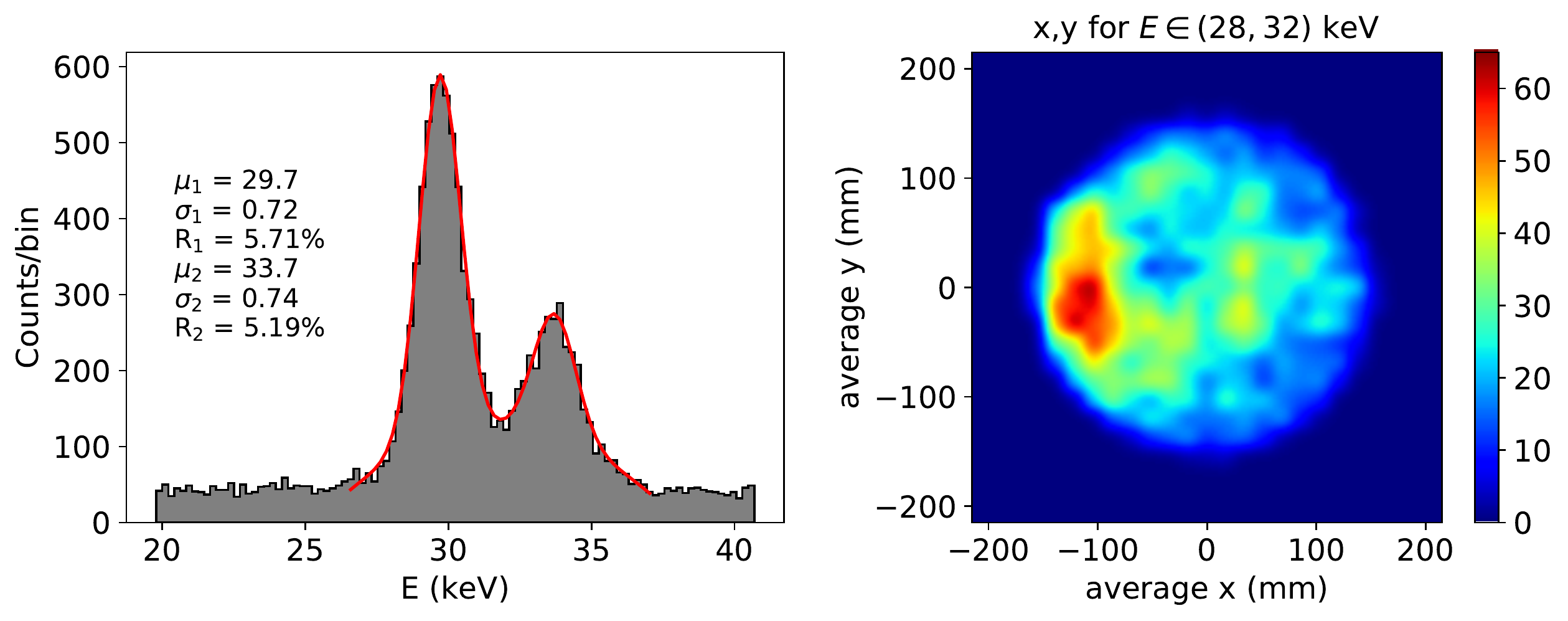}
	\caption{(Left) Fit of the xenon x-ray peaks to two Gaussians + a second-order polynomial.  (Right) The resulting $(x,y)$ distribution of events included in the fit.} \label{fig.epeaks.exrays}
\end{figure}

\Fig\ \ref{fig.epeaks.exrays} shows the energy distribution of isolated low-energy clusters identified inside the fiducial volume.  Two visible peaks are attributed to several lines near \XenonXRaysLineAlphaEnergy\ and several lines near \XenonXRaysLineBetaEnergy. The observed distribution is well described by a combination of two Gaussians and a second degree polynomial.  As the more intense X-ray lines in the lower-energy peak are much closer together than in the higher-energy peak, the width of the leftmost Gaussian, \ResolutionXRays\ FWHM, is taken to represent the energy resolution at the energy of \XenonXRaysLineAlphaEnergy. The error on the resolution is mostly of systematic origin and it was estimated by varying the binning and fit ranges, slightly varying the fiducial cuts, and considering the statistical errors of the fit, which were less than or equal to half of the stated error.

\subsection*{Higher-energy gammas}

\begin{figure}[htb]
	\centering
	\includegraphics[width= 1.0\textwidth]{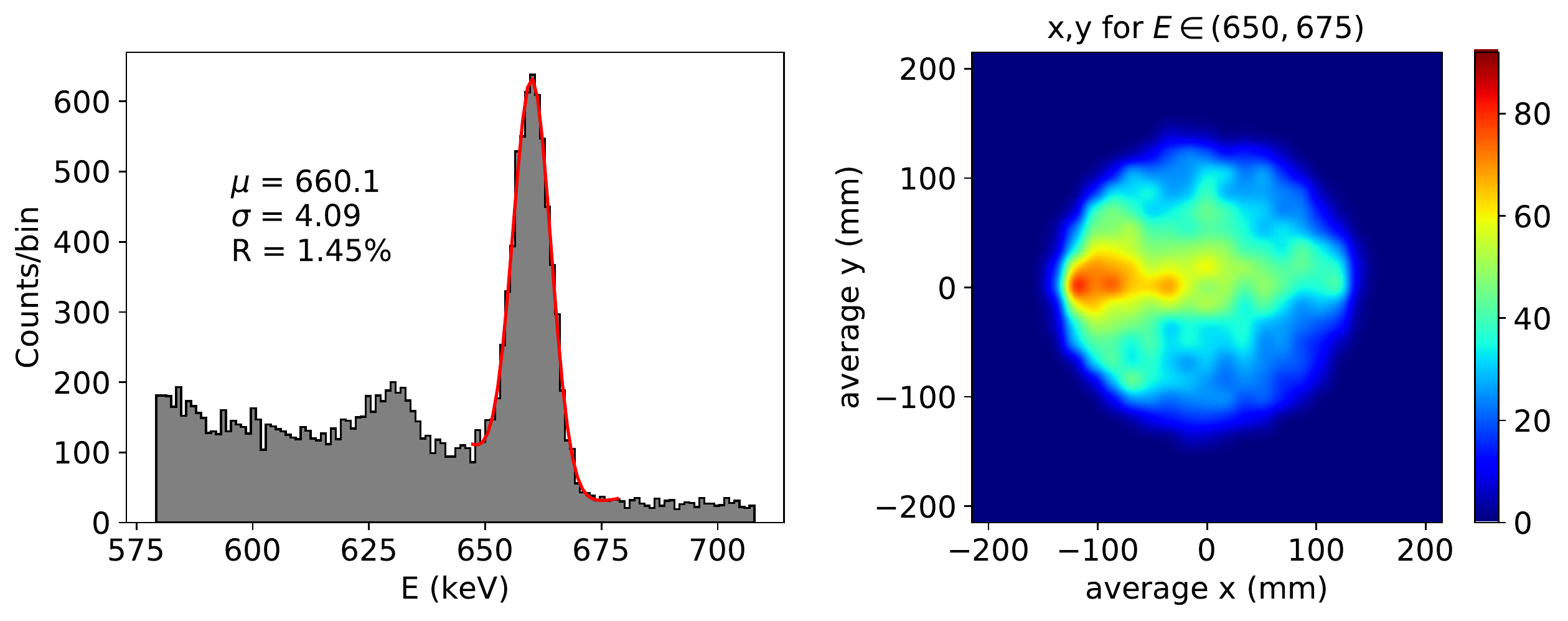}
	\includegraphics[width= 1.0\textwidth]{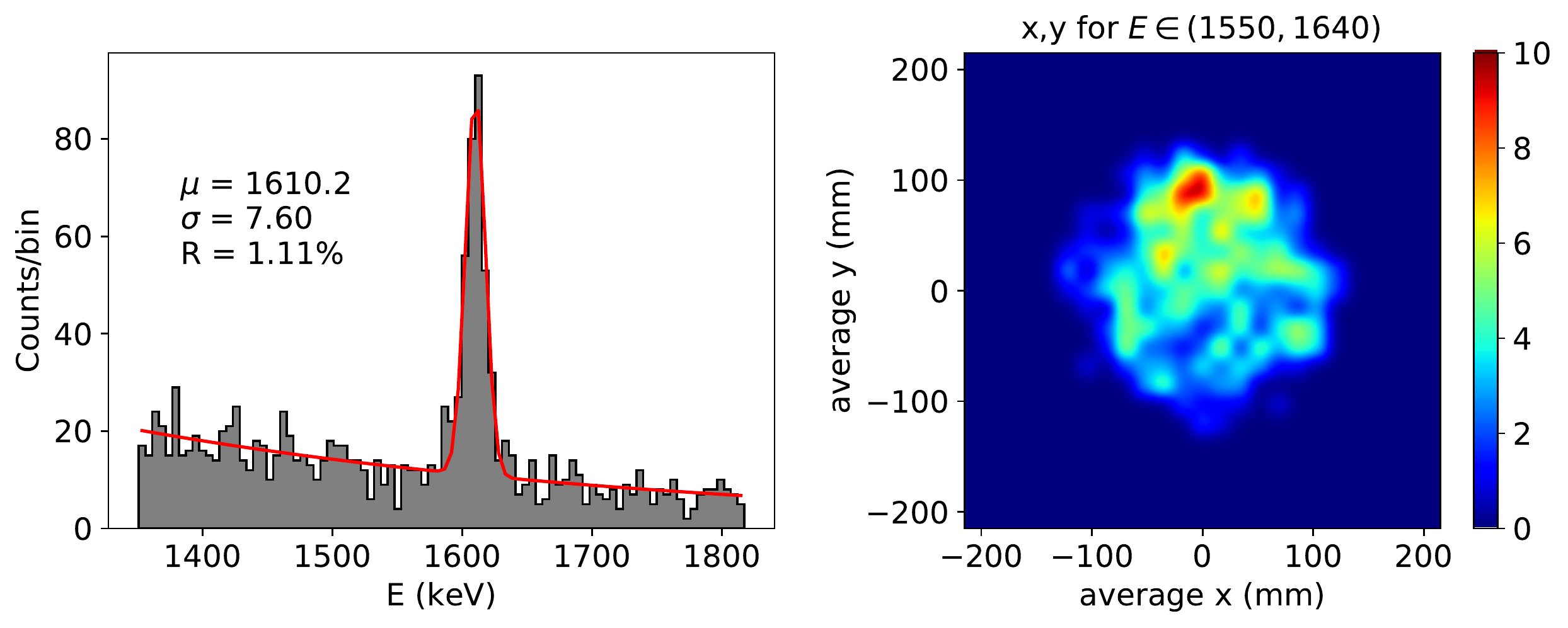}
	\caption{(Top left) Fit of the \CS\ photo-peak to a Gaussian + second-order polynomial in the selected optimal fiducial region and (top right) the resulting $(x,y)$ distribution of events included in the fit.  (Bottom left) Fit of the \TlDoubleEscapePeakEnergy\ double-escape peak in the selected optimal fiducial region and (bottom right) the resulting $(x,y)$ distribution of events in the peak region.} \label{fig.ecsth}
\end{figure}

To determine the detector's energy resolution at higher energies, the observed energy distribution in the neighborhood of the \CsGammaEnergy\ peak was fitted as the sum of a Gaussian + 2$^{\mathrm{nd}}$ order polynomial, and, in a similar fashion, the distribution of the observed energy in the region of  the double-escape peak was fitted as the sum of a Gaussian + exponential (see \Fig\ \ref{fig.ecsth}). The fits yield the results: $R(\CsGammaEnergy) = \ResolutionCsRF$ FWHM and $R(\TlDoubleEscapePeakEnergy) = \ResolutionTlRF$ FWHM. The errors on the resolution were estimated in the same way as for the X-ray case.

 \begin{figure}[htb]
 	\centering
 	\includegraphics[width= 1.0\textwidth]{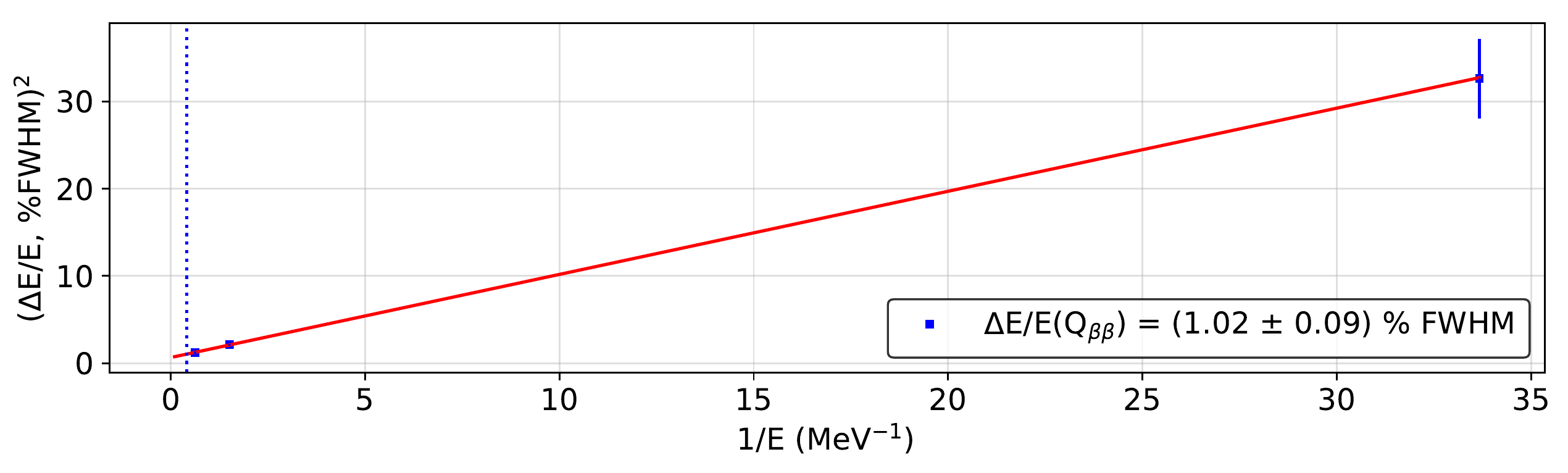}
 	\caption{Square of the resolution (\%FWHM)$^2$ as a function of $1/E$. The red line represents a fit to the \Eq\ \ref{eq.resolution}, the dotted blue vertical line indicates the position of \Qbb.} \label{fig.extrap}
 \end{figure}

By performing a fit to \Eq\ \ref{eq.resolution}, the measured values of the energy resolution can be extrapolated to \Qbb\ yielding $R(\Qbb) = (1.02 \pm 0.09) \%$ FWHM.  The fit, shown in \Fig\ \ref{fig.extrap}, gives $a = (0.98 \pm 0.07)\%$ FWHM $\cdot$ MeV and indicates the presence of a significant constant term $c = (0.80 \pm 0.11)\%$ FWHM which dominates the energy resolution at \Qbb. This term reflects the initial state of the understanding of the detector and it is likely related to a relatively poor electron lifetime during the calibration runs.

\section{Summary}
\label{sec.conclu}

Nearly monoenergetic gamma rays with energies ranging from \XenonXRaysLineAlphaEnergy\ (xenon \XenonKAlpha\ x-ray line) to 
\TlDoubleEscapePeakEnergy\ (\TL\ double-escape peak) have been used to study the performance of the \HPXe\ detector \NEW, operating at the LSC in Canfranc. The energy reconstruction and detailed response calibration procedures described in \cite{Martinez-Lema:2018ibw} were applied to data taken using \CS\ and \THO\ calibration sources and demonstrate that the response of the \NEW\ detector is linear to within approximately 1\% over the energy range studied. The attained energy resolution varies from \ResolutionXRays\ FWHM at \XenonXRaysLineAlphaEnergy\ to \ResolutionTlRF\ FWHM at \TlDoubleEscapePeakEnergy .

The assumption that the energy resolution is a combination of a stochastic term and a constant term leads to the conclusion that the energy resolution of the present detector at \Qbb\ is of the order of $(1.02 \pm 0.09) \%$.  Thus energy resolution is dominated by systematic effects reflecting the initial understanding of the performance of the detector and the relatively poor electron lifetime present during the calibration runs, and so a significant improvement of the energy resolution can be expected in the future.  Note that this energy resolution was obtained using the ionization signal only. Expressed in terms of $\sigma/E$, the \NEW\ detector has achieved the resolution of $0.43\%$ at \Qbb, which compares favorably with the best energy resolution obtained using a combination of ionization and scintillation signals in liquid xenon in Phase II of the EXO experiment \cite{PhysRevLett.120.072701} of $\sigma/E = 1.23\%$.

\acknowledgments
The NEXT Collaboration acknowledges support from the following agencies and institutions: the European Research Council (ERC) under the Advanced Grant 339787-NEXT; the European Union's Framework Programme for Research and Innovation Horizon 2020 (2014-2020) under the Marie Sk\l{}odowska-Curie Grant Agreements No. 674896, 690575 and 740055; the Ministerio de Econom\'ia y Competitividad of Spain under grants FIS2014-53371-C04, the Severo Ochoa Program SEV-2014-0398 and the Mar\'ia de Maetzu Program MDM-2016-0692; the GVA of Spain under grants PROMETEO/2016/120 and SEJI/2017/011; the Portuguese FCT and FEDER through the program COMPETE, projects PTDC/FIS-NUC/2525/2014 and UID/FIS/04559/2013; the U.S.\ Department of Energy under contract numbers DE-AC02-07CH11359 (Fermi National Accelerator Laboratory), DE-FG02-13ER42020 (Texas A\&M), DE-SC0017721 (University of Texas at Arlington), and DE-AC02-06CH11357 (Argonne National Laboratory); and the University of Texas at Arlington. We also warmly acknowledge the Laboratorio Nazionale di Gran Sasso (LNGS) and the Dark Side collaboration for their help with TPB coating of various parts of the NEXT-White TPC. Finally, we are grateful to the Laboratorio Subterr\'aneo de Canfranc for hosting and supporting the NEXT experiment.

\bibliographystyle{pool/JHEP}
\bibliography{pool/NextRefs}

\providecommand{\href}[2]{#2}\begingroup\raggedright\begin{thebibliography}{10}

\bibitem{BOLOTNIKOV1997360}
A.~Bolotnikov and B.~Ramsey, {\it The spectroscopic properties of high-pressure
  xenon},  {\em Nucl. Instrum. Meth.} {\bf A396} (1997) 360.

\bibitem{Nygren:2009zz}
D.~Nygren, {\it {High-pressure xenon gas electroluminescent TPC for 0-$\nu ~
  \beta \beta$-decay search}},  {\em Nucl. Instrum. Meth.} {\bf A603} (2009)
  337.

\bibitem{Alvarez:2011my}
{\bf NEXT} Collaboration, V.~\'Alvarez et~al., {\it {The NEXT-100 experiment
  for neutrinoless double beta decay searches (Conceptual Design Report)}},
  \href{http://xxx.lanl.gov/abs/1106.3630}{{\tt arXiv:1106.3630}}.

\bibitem{Alvarez:2012haa}
{\bf NEXT} Collaboration, V.~\'Alvarez et~al., {\it {NEXT-100 Technical Design
  Report (TDR): Executive Summary}},  {\em JINST} {\bf 7} (2012) T06001,
  [\href{http://xxx.lanl.gov/abs/1202.0721}{{\tt arXiv:1202.0721}}].

\bibitem{Gomez-Cadenas:2013lta}
{\bf NEXT} Collaboration, J.~J. Gomez-Cadenas et~al., {\it {Present status and
  future perspectives of the NEXT experiment}},  {\em Adv. High Energy Phys.}
  {\bf 2014} (2014) 907067, [\href{http://xxx.lanl.gov/abs/1307.3914}{{\tt
  arXiv:1307.3914}}].

\bibitem{Martin-Albo:2015rhw}
{\bf NEXT} Collaboration, J.~Mart\'in-Albo et~al., {\it {Sensitivity of
  NEXT-100 to Neutrinoless Double Beta Decay}},  {\em JHEP} {\bf 05} (2016)
  159, [\href{http://xxx.lanl.gov/abs/1511.09246}{{\tt arXiv:1511.09246}}].

\bibitem{Alvarez:2012hh}
{\bf NEXT} Collaboration, V.~\'Alvarez et~al., {\it {Near-Intrinsic Energy
  Resolution for 30 to 662 keV Gamma Rays in a High Pressure Xenon
  Electroluminescent TPC}},  {\em Nucl.\ Instrum.\ Meth.} {\bf A708} (2012)
  101, [\href{http://xxx.lanl.gov/abs/1211.4474}{{\tt arXiv:1211.4474}}].

\bibitem{Alvarez:2012xda}
{\bf NEXT} Collaboration, V.~\'Alvarez et~al., {\it {Initial results of
  NEXT-DEMO, a large-scale prototype of the NEXT-100 experiment}},  {\em JINST}
  {\bf 8} (2013) P04002, [\href{http://xxx.lanl.gov/abs/1211.4838}{{\tt
  arXiv:1211.4838}}].

\bibitem{Monrabal:2018xlr}
{\bf NEXT} Collaboration, F.~Monrabal et~al., {\it {The Next White (NEW)
  Detector}},  \href{http://xxx.lanl.gov/abs/1804.02409}{{\tt
  arXiv:1804.02409}}.

\bibitem{Martinez-Lema:2018ibw}
{\bf NEXT} Collaboration, G.~Mart\'inez-Lema et~al., {\it {Calibration of the
  NEXT-White detector using $^{83m}\mathrm{Kr}$ decays}},
  \href{http://xxx.lanl.gov/abs/1804.01780}{{\tt arXiv:1804.01780}}.

\bibitem{Simon:2018vep}
{\bf NEXT} Collaboration, A.~Sim\'on et~al., {\it {Electron drift properties in
  high pressure gaseous xenon}},  {\em JINST} {\bf 13} (2018) P07013,
  [\href{http://xxx.lanl.gov/abs/1804.01680}{{\tt arXiv:1804.01680}}].

\bibitem{LBLTOI}
L.~P. Ekstr{\"o}m and R.~B. Firestone, {\em {WWW Table of Radioactive Isotopes,
  database version 2/28/99}}.
\newblock
  \texttt{\href{http://nucleardata.nuclear.lu.se/toi}{http://nucleardata.nuclear.lu.se/toi}}.

\bibitem{PhysRevLett.120.072701}
{\bf EXO} Collaboration, J.~B. Albert et~al., {\it Search for neutrinoless
  double-beta decay with the upgraded {EXO}-200 detector},  {\em Phys. Rev.
  Lett.} {\bf 120} (2018) 072701,
  [\href{http://xxx.lanl.gov/abs/1707.08707}{{\tt arXiv:1707.08707}}].

\end{thebibliography}\endgroup

\end{document}